\documentclass[aps,pra,preprint,groupedaddress, superscriptaddress, citesort]{revtex4}

\usepackage{amssymb,}
\usepackage{amsmath}
\usepackage{graphics}
\usepackage{graphicx}
\usepackage{float}
\def\be{\begin{equation}}
\def\ee{\end{equation}}
\def\ba{\begin{array}}
\def\ea{\end{array}}

\begin{document}

\title{One-way deficit of $SU(2)$ invariant states}
\author{Yao-Kun Wang}
\affiliation{School of Mathematical Sciences,  Capital Normal
University,  Beijing 100048,  China}
\affiliation{College of Mathematics,  Tonghua Normal University, Tonghua 134001,  China}
\author{Teng Ma}
\affiliation{School of Mathematical Sciences,  Capital Normal University,  Beijing 100048,  China}
\author{Shao-Ming Fei}
\affiliation{School of Mathematical Sciences,  Capital Normal University,  Beijing 100048,  China}
\affiliation{Max-Planck-Institute for Mathematics in the Sciences, 04103 Leipzig, Germany}
\author{Zhi-Xi Wang}
\affiliation{School of Mathematical Sciences,  Capital Normal
University,  Beijing 100048,  China}

\begin{abstract}
We calculate analytically the one-way deficit for systems
consisting spin-$j$ and spin-$\frac{1}{2}$ subsystems with
$SU(2)$ symmetry. Comparing our results with the quantum
discord of $SU(2)$ invariant states, we show that the one-way deficit is equal to the quantum
discord for half-integer $j$, and is larger than the quantum
discord for integer $j$. Moreover, we also compare the one-way
deficit with entanglement of formation. The quantum entanglement tends to zero as $j$ increases,
while the one-way deficit can remain significantly large.
\end{abstract}
\maketitle

\section{Introduction}

Quantum entanglement has played significant roles in the field of
quantum information and quantum computation such as super-dense
coding \cite{CH}, teleportation \cite{CH2},  quantum
cryptography \cite{AK}, remote-state preparation \cite{AK2}.
Quantum correlations other than quantum entanglement have also
attracted much attention recently \cite{bennett,
zurek1,1modi,Auccaise,Giorgi,1Streltsov,modi2}. Among of them,
quantum discord introduced by Oliver and Zurek and independently by
Henderson and Vedral \cite{zurek1} is a widely
accepted quantity. Quantum discord, a measure which quantifies the
discrepancy between the quantum mutual information and the maximal classical
information, can be present in separable mixed quantum
states. Following this discovery, much work has been done in
investigating the properties and behavior of quantum discord in
various systems. Since complicated optimization procedure
involved in calculating quantum discord, one has no analytical formulae of quantum discord
even for two-qubit quantum systems \cite{luo,Ali,Li,chen,shi,Vinjanampathy}.

Other significant nonclassical correlations besides entanglement and
quantum discord, for example, the quantum
deficit \cite{oppenheim,horodecki}, measurement-induced disturbance
\cite{luo}, symmetric discord \cite{piani,wu}, relative entropy of
discord and dissonance \cite{modi}, geometric discord
\cite{luoandfu,dakic}, and continuous-variable discord
\cite{adesso,giorda} have been studied recently.
The work deficit \cite{oppenheim} proposed  to characterize quantum
correlations in terms of entropy  production and work extraction by
Oppenheim \emph{et al } is the first operational approach to connect the
quantum correlations theory with quantum thermodynamics. Recently,
Alexander Streltsov \emph{et al }\cite{Streltsov0,chuan} reveals
that the one-way deficit plays an important role in quantum correlations
as a resource for the distribution of entanglement. The one-way deficit
of a bipartite quantum state $\rho$ is defined by \cite{horodecki2}:
\begin{eqnarray}
\Delta^{\rightarrow}(\rho)=\min\limits_{\{\Pi_{k}\}}S(\sum\limits_{i}\Pi_{k}\rho\Pi_{k})-S(\rho),\label{deficitdefinition}
\end{eqnarray}
where $\{\Pi_{k}\}$ is the projective measurements and $S$ is the von Neumann entropy.

The one-way deficit and quantum discord have similar minimum form but they are
different kinds of quantum correlations. We
have obtained analytical formula of one-way deficit for some
well known states such as Bell-diagonal states \cite{wang}.
In this paper, we endeavored to calculate the one-way deficit of
bipartite $SU(2)$ invariant states consisting of a spin-$j$
and a spin-$\frac{1}{2}$ subsystems.

$SU(2)$-invariant density matrices of two spins $\vec S_{1}$ and $\vec S_{2}$
are defined to be invariant under $U_{1}\otimes U_{2}$,  $U_1\otimes U_2\rho
U^{\dagger}_1\otimes U^{\dagger}_2=\rho$, where
$U_{a}=\exp(i\vec\eta\cdot\vec S_{a})$, $a\in\{1,2\}$, are the usual
rotation operator representation of $SU(2)$ with real parameter
$\vec\eta$ and $\hbar=1$ \cite{Schliemann1,Schliemann2}. Those $SU(2)$-invariant
states $\rho$ commute with all the components of the total spin $\vec
J=\vec S_{1}+\vec S_{2}$. In real physical systems,
$SU(2)$-invariant states arise from thermal equilibrium
states of spin systems described by $SU(2)$ invariant
Hamiltonian \cite{Durkin}. The state space structure and entanglement of
SO(3)-invariant bipartite quantum systems have been analyzed in
the literature \cite{Breuer1,Breuer2}. For $SU(2)$ invariant quantum
spin systems, negativity is shown to be necessary and
sufficient for separability \cite{Schliemann1,Schliemann2},
and the relative entropy of entanglement has been analytically
calculated \cite{zwang} for $(2j+1)\otimes 2$ and $(2j+1)\otimes 3$
dimensional systems. Furthermore, the entanglement of formation (EoF), I-concurrence, I-tangle
and convex-roof-extended negativity of the $SU(2)$-invariant states
of a spin-$j$ and spin-$\frac{1}{2}$ \cite{Manne} have been
analytically calculated by using the approach in
\cite{Vollbrecht}. Quantum discord for $SU(2)$-invariant states
composed of spin-$j$ and spin-$\frac{1}{2}$ systems has been analytically calculated in \cite{cakmak}.

As an $SU(2)$-invariant
state commutes with all the components of $\vec J$, $\rho$ has the general form \cite{Schliemann1},
\begin{equation}
\rho=\sum_{J=|S_{1}-S_{2}|}^{S_{1}+S_{2}}
\frac{A(J)}{2J+1}\sum_{J^{z}=-J}^{J}
|J,J^{z}\rangle_{0}{_{0}\langle J,J^{z}|}\,,
\end{equation}
where the constants $A(J)\geq 0$, $\sum_{J}A(J)=1$,
$|J,J^{z}\rangle_{0}$ denotes a state of total spin $J$ and
$z$-component $J^{z}$. We consider the case with $\vec S_{1}$ of arbitrary length $S$ and
$\vec S_{2}$ of length $\frac{1}{2}$. Let $S=j$,
$J^{z}=m$. A general $SU(2)$-invariant density matrix has the form,
\begin{eqnarray}
\rho^{ab}=\frac{F}{2j}\sum_{m=-j+\frac{1}{2}}^{j-\frac{1}{2}}|j-\frac{1}{2}, m\rangle
\langle j-\frac{1}{2}, m|+\frac{1-F}{2(j+1)}\sum_{m=-j-\frac{1}{2}}^{j+\frac{1}{2}}|j
+\frac{1}{2}, m\rangle\langle j+\frac{1}{2}, m|, \label{1state}
\end{eqnarray}
where $F\in[0,1]$, and $F$ is a function of temperature in thermal
equilibrium. $\rho^{ab}$ is a $(2j+1)\otimes 2$ bipartite state. It has two eigenvalues
$\lambda_1={F}/{(2j)}$ and $\lambda_2=({1-F})/({2j+2})$ with
degeneracies $2j$ and $2j+2$, respectively \cite{cakmak}. The entropy
of  $\rho^{ab}$ is given by
\begin{eqnarray}
S(\rho^{ab})=-F\log (\frac{F}{2j})-(1-F)\log (\frac{1-F}{2j+2}).\label{entropy}
\end{eqnarray}
As the eigenstates of the total spin can be given by the
Clebsch-Gordon coefficients \cite{shankar} in coupling a spin-$j$
to spin-$\frac{1}{2}$,
\begin{eqnarray}
|j\pm\frac{1}{2},m\rangle=
\pm\sqrt{\frac{j+\frac{1}{2}\pm m}{2j+1}}
|j,m-\frac{1}{2}\rangle\otimes
|\frac{1}{2},\frac{1}{2}\rangle
+\sqrt{\frac{j+\frac{1}{2}\mp m}{2j+1}}
|j,m+\frac{1}{2}\rangle\otimes
|\frac{1}{2},-\frac{1}{2}\rangle,
\end{eqnarray}
the density matrix (\ref{1state}) can be written in product basis form \cite{cakmak},
\begin{eqnarray}
\rho^{ab}&=&\frac{F}{2j}\sum_{m=-j+\frac{1}{2}}^{j-\frac{1}{2}}
(a_-^2|m-\frac{1}{2}\rangle\langle m-\frac{1}{2}|\otimes
|\frac{1}{2}\rangle\langle \frac{1}{2}| \\\nonumber
& &+a_-b_-(|m-\frac{1}{2}\rangle\langle m+\frac{1}{2}|\otimes |\frac{1}{2}\rangle\langle -\frac{1}{2}| \\\nonumber
& &+|m+\frac{1}{2}\rangle\langle m-\frac{1}{2}|\otimes |-\frac{1}{2}\rangle\langle \frac{1}{2}|) \\\nonumber
& &+b_-^2|m+\frac{1}{2}\rangle\langle m+\frac{1}{2}|\otimes |-\frac{1}{2}\rangle\langle -\frac{1}{2}|)\\\nonumber
& & +\frac{1-F}{2(j+1)}\sum_{m=-j-\frac{1}{2}}^{j+\frac{1}{2}}(a_+^2|m-\frac{1}{2}\rangle\langle m-\frac{1}{2}|\otimes |\frac{1}{2}\rangle\langle \frac{1}{2}| \\\nonumber
& &+a_+b_+(|m-\frac{1}{2}\rangle\langle m+\frac{1}{2}|\otimes |\frac{1}{2}\rangle\langle -\frac{1}{2}| \\\nonumber
& &+|m+\frac{1}{2}\rangle\langle m-\frac{1}{2}|\otimes |-\frac{1}{2}\rangle\langle \frac{1}{2}|) \\\nonumber
& &+b_+^2|m+\frac{1}{2}\rangle\langle m+\frac{1}{2}|\otimes |-\frac{1}{2}\rangle\langle -\frac{1}{2}|),\nonumber
\end{eqnarray}
where $a_{\pm}=\pm\sqrt{\frac{j+\frac{1}{2}\pm m}{2j+1}}$ and
$b_{\pm}=\sqrt{\frac{j+\frac{1}{2}\mp m}{2j+1}}$.

When $j=\frac{1}{2}$, the state $\rho^{ab}$ turns out to be the $2\otimes2$ Werner state:
\begin{eqnarray}
\rho=(1-c)\frac{I}{4}+c|\psi^{-}\rangle\langle\psi^{-}|,~~~ c=\frac{4F-1}{3},
\end{eqnarray}
with $|\psi^{-}\rangle=\frac{1}{\sqrt{2}}(|01\rangle-|10\rangle)$.

\section{Main result}

Any von Neumann measurement on the spin-$\frac{1}{2}$ subsystem can
be written as: 
\begin{eqnarray}
B_k=V\Pi_kV^{\dag},~~~~k=0,1,
\end{eqnarray}
where $\Pi_k=|k\rangle\langle k|$, $|k\rangle$ is the computational basis,
$V=tI+i\vec{y}\cdot\vec{\sigma}\in SU(2)$, $\vec{\sigma}=(\sigma_1,\sigma_2,\sigma_3)$ with $\sigma_1,\sigma_2,\sigma_3$ being Pauli matrices,
$t\in \Re$ and $\vec{y}=(y_{1},y_{2},y_{3})\in \Re^{3}$, $t^2+y_1^2+y_2^2+y_3^2=1$.
Set
\begin{eqnarray*}
M=\sum_{m=-j}^{j}& &(z_3\frac{m(2Fj+F-j)}{j(j+1)(2j+1)}|m\rangle\langle
m|\\
& &+(z_1+iz_2)\frac{\sqrt{j(j+1)-m(m+1)}(2Fj+F-j)}{2j(j+1)(2j+1)}|m\rangle\langle
m+1|  \\
& &+(z_1-iz_2)\frac{\sqrt{j(j+1)-m(m+1)}(2Fj+F-j)}{2j(j+1)(2j+1)}|m+1\rangle\langle
m|).
\end{eqnarray*}
where $z_1=2(-ty_2+y_1y_3)$, $z_2=2(ty_1+y_2y_3)$, $z_3=t^2+y_3^2-y_1^2-y_2^2$ with $z_1^2+z_2^2+z_3^2=1$.
After the measurement, one has the ensemble of
post-measurement states $\{\rho_k, p_k\}$ with $p_0=p_1=\frac{1}{2}$
and the corresponding post-measurement states \cite{cakmak},
\begin{eqnarray} \label{r1}
\rho_0&=&(\frac{1}{2j+1}\sum_{m=-j}^{j}|m\rangle\langle
m|-M)\otimes V\Pi_0V^{\dag}\\ \nonumber
&=&(\frac{1}{2j+1}I-M)\otimes V\Pi_0V^{\dag},
\end{eqnarray}
and
\begin{eqnarray}\label{r2}
\rho_1&=&(\frac{1}{2j+1}\sum_{m=-j}^{j}|m\rangle\langle
m|+M)\otimes V\Pi_1V^{\dag}\\ \nonumber
&=&(\frac{1}{2j+1}I+M)\otimes V\Pi_1V^{\dag},
\end{eqnarray}

The eigenvalues of $\rho_0$, $\rho_1$ \cite{cakmak} are:
\begin{eqnarray}
\lambda_n^{\pm}=\frac{1}{2j+1}\pm\frac{j-n}{j(j+1)(2j+1)}\lvert (F(2j+1)-j)\rvert,\label{eigenvalue1}
\end{eqnarray}
where $n=0,\cdots, \lfloor j\rfloor$, and $\lfloor j \rfloor$ denotes the largest integer that is less or equal to $j$.
Obviously the eigenvalues are independent of the measurement. Due to this fact, analytical expression
for quantum discord of $\rho^{ab}$ has been obtained in \cite{cakmak},
\begin{eqnarray}
D(\rho^{ab})&=&F\text{log}(\frac{F}{2j})+(1-F)\text{log}(\frac{1-F}{2j+2})+1\\ \nonumber
& &-\sum_{n=0}^{\lfloor j\rfloor}(\frac{1}{2j+1}+\frac{j-n}{j(j+1)(2j+1)}\lvert (F(2j+1)-j)\rvert)\\ \nonumber
& &\cdot\text{log}(\frac{1}{2j+1}+\frac{j-n}{j(j+1)(2j+1)}\lvert (F(2j+1)-j)\rvert)\\ \nonumber
& &-\sum_{n=0}^{\lfloor j\rfloor}(\frac{1}{2j+1}-\frac{j-n}{j(j+1)(2j+1)}\lvert (F(2j+1)-j)\rvert)\\ \nonumber
& &\cdot\text{log}(\frac{1}{2j+1}-\frac{j-n}{j(j+1)(2j+1)}\lvert (F(2j+1)-j)\rvert).
\end{eqnarray}

To evaluate the one-way deficit of $\rho^{ab}$, we calculate the eigenvalues  of
$\sum\limits_{i}\Pi_{k}\rho^{ab}\Pi_{k}=p_{0}\rho_{0}+p_{1}\rho_{1}$.
From (\ref{r1}) and (\ref{r2}), since $\frac{1}{2j+1}I-M$ commutes
with $\frac{1}{2j+1}I+M$, by using (19) and (20) in \cite{wang}, we have the
eigenvalues of $\sum\limits_{i}\Pi_{k}\rho^{ab}\Pi_{k}$,
\begin{eqnarray}
\ \mu_n^{\pm}=\frac{1}{2}\lambda_n^{\pm},
\end{eqnarray}
with each algebraic multiplicity two. As the
eigenvalues do not depend on the measurement  parameters, the minimum of the entropy of the
post-measurement states do not require any optimization over the
projective measurements,
\begin{eqnarray}
\min\limits_{\{\Pi_{k}\}}S(\sum_{i}\Pi_{k}\rho^{ab}\Pi_{k})&=&-\sum_{n=0}^{\lfloor j\rfloor}(\frac{1}{(2j+1)}\pm\frac{j-n}{j(j+1)(2j+1)}\lvert (F(2j+1)-j)\rvert)\nonumber\\
& &\cdot\log(\frac{1}{2(2j+1)}\pm\frac{j-n}{2j(j+1)(2j+1)}\lvert (F(2j+1)-j)\rvert). \label{min1}
\end{eqnarray}
From Eqs. (\ref{deficitdefinition}), (\ref{entropy}) and (\ref{min1}), the one-way deficit of the state is given by
\begin{eqnarray}\label{deficit}
\Delta^{\rightarrow}(\rho^{ab})&=&\min\limits_{\{\Pi_{k}\}}S(\sum\limits_{i}\Pi_{k}\rho^{ab}\Pi_{k})-S(\rho^{ab})\nonumber\\
&=&F\log (\frac{F}{2j})+(1-F)\log (\frac{1-F}{2j+2})\nonumber\\
& &-\sum_{n=0}^{\lfloor j\rfloor}(\frac{1}{(2j+1)}+\frac{j-n}{j(j+1)(2j+1)}\lvert (F(2j+1)-j)\rvert)\nonumber\\
& &\cdot\log\frac{1}{2}(\frac{1}{(2j+1)}+\frac{j-n}{j(j+1)(2j+1)}\lvert (F(2j+1)-j)\rvert)\nonumber\\
& &-\sum_{n=0}^{\lfloor j\rfloor}(\frac{1}{(2j+1)}-\frac{j-n}{j(j+1)(2j+1)}\lvert (F(2j+1)-j)\rvert)\nonumber\\
& &\cdot\log\frac{1}{2}(\frac{1}{(2j+1)}-\frac{j-n}{j(j+1)(2j+1)}\lvert (F(2j+1)-j)\rvert)\nonumber\\
&=&F\text{log}(\frac{F}{2j})+(1-F)\text{log}(\frac{1-F}{2j+2})+\frac{2}{2j+1}(\lfloor j\rfloor +1)\\ \nonumber
& &-\sum_{n=0}^{\lfloor j\rfloor}(\frac{1}{2j+1}+\frac{j-n}{j(j+1)(2j+1)}\lvert (F(2j+1)-j)\rvert)\\ \nonumber
& &\cdot\text{log}(\frac{1}{2j+1}+\frac{j-n}{j(j+1)(2j+1)}\lvert (F(2j+1)-j)\rvert)\\ \nonumber
& &-\sum_{n=0}^{\lfloor j\rfloor}(\frac{1}{2j+1}-\frac{j-n}{j(j+1)(2j+1)}\lvert (F(2j+1)-j)\rvert)\\ \nonumber
& &\cdot\text{log}(\frac{1}{2j+1}-\frac{j-n}{j(j+1)(2j+1)}\lvert (F(2j+1)-j)\rvert)\\ \nonumber
&=&\left\{ \begin{array}{ll}
        D(\rho^{ab}), & j\ \  \text{is a half-integer} \\
        D(\rho^{ab})+\frac{1}{d}, &j\ \   \text{is an integer}, \end{array}\right.
\end{eqnarray}
where $d=2j+1$. Especially, when $j=\frac{1}{2}$, the state becomes the
$2\otimes2$ Werner state, and the one-way deficit is equal to the
quantum discord, which is in consistent with the result obtained in \cite{wang}.

\begin{figure}[h]
\raisebox{12em}{(a)}\includegraphics[width=6.25cm]{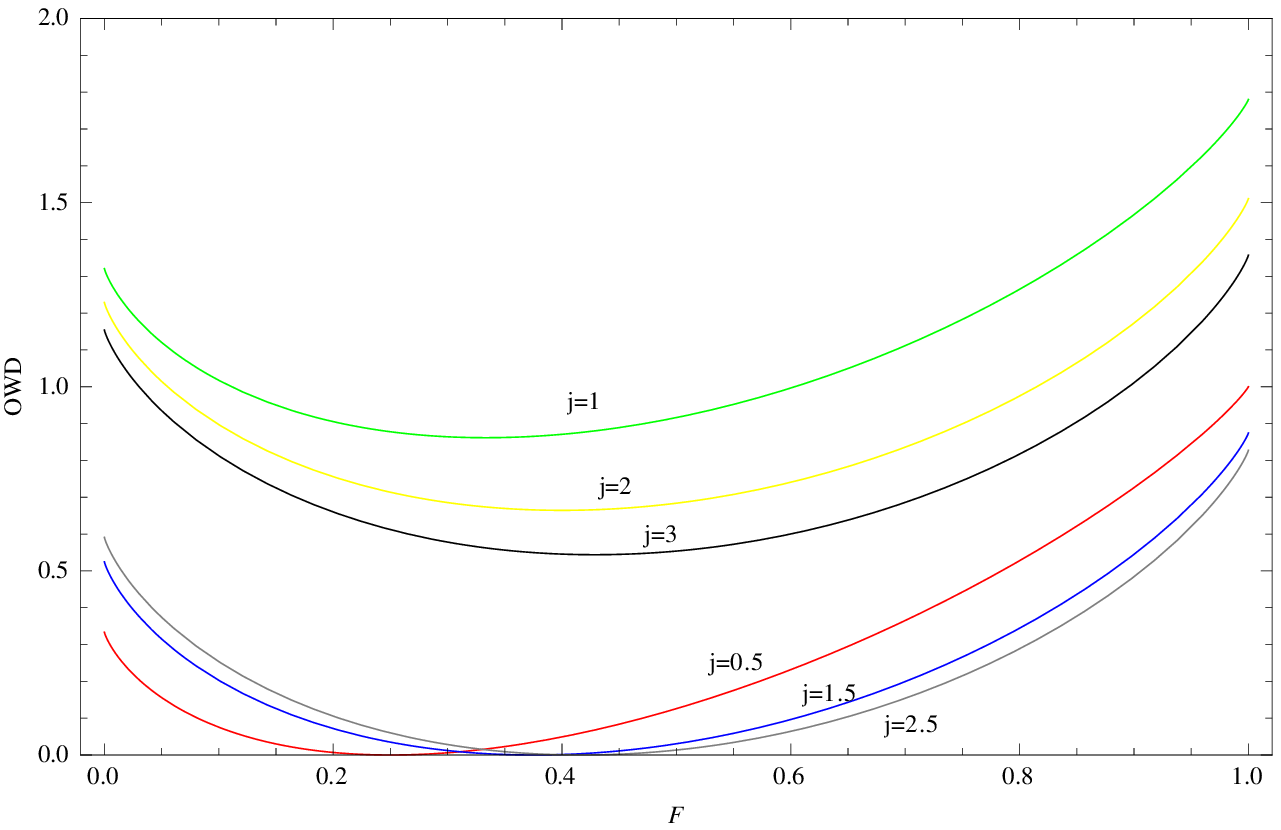}
\qquad
\raisebox{12em}{(b)}\includegraphics[width=6.25cm]{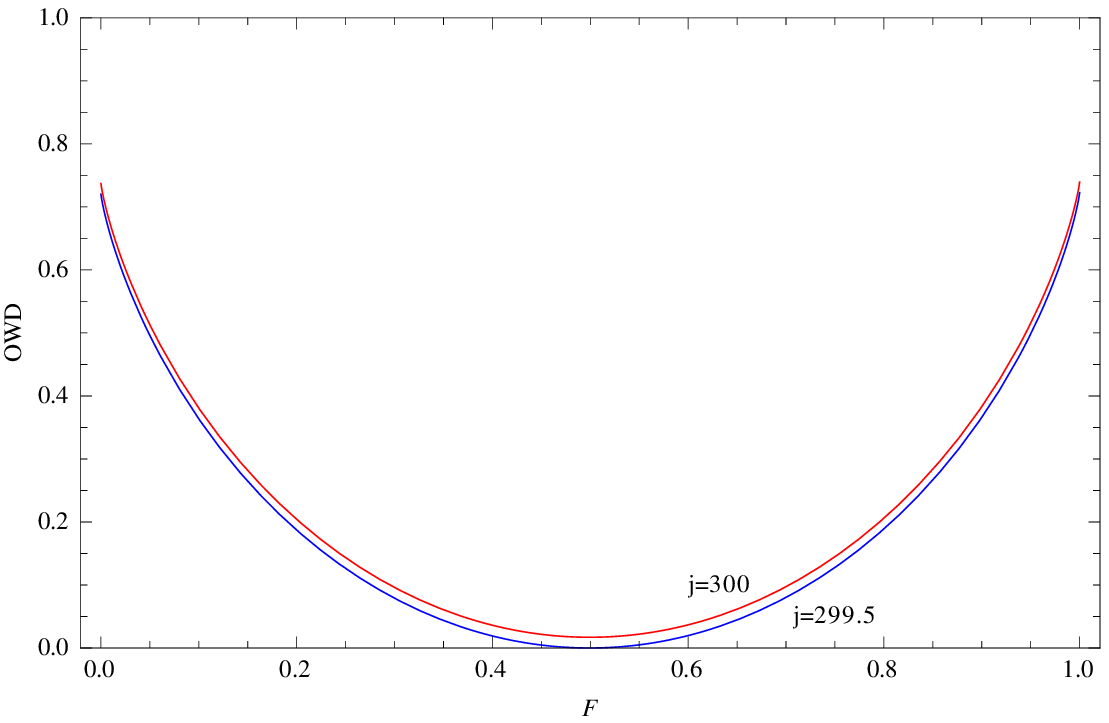}
\begin{center}
\raisebox{12em}{(c)}\includegraphics[width=6.25cm]{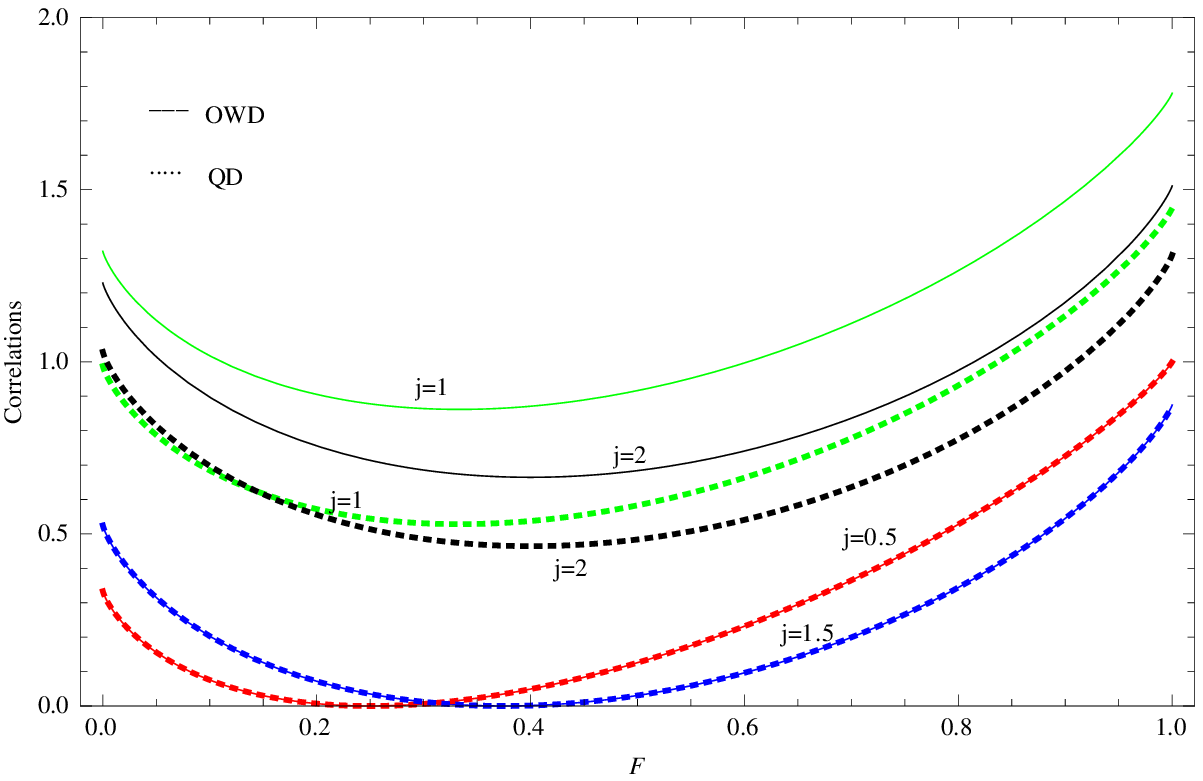}
\qquad
\raisebox{12em}{(d)}\includegraphics[width=6.25cm]{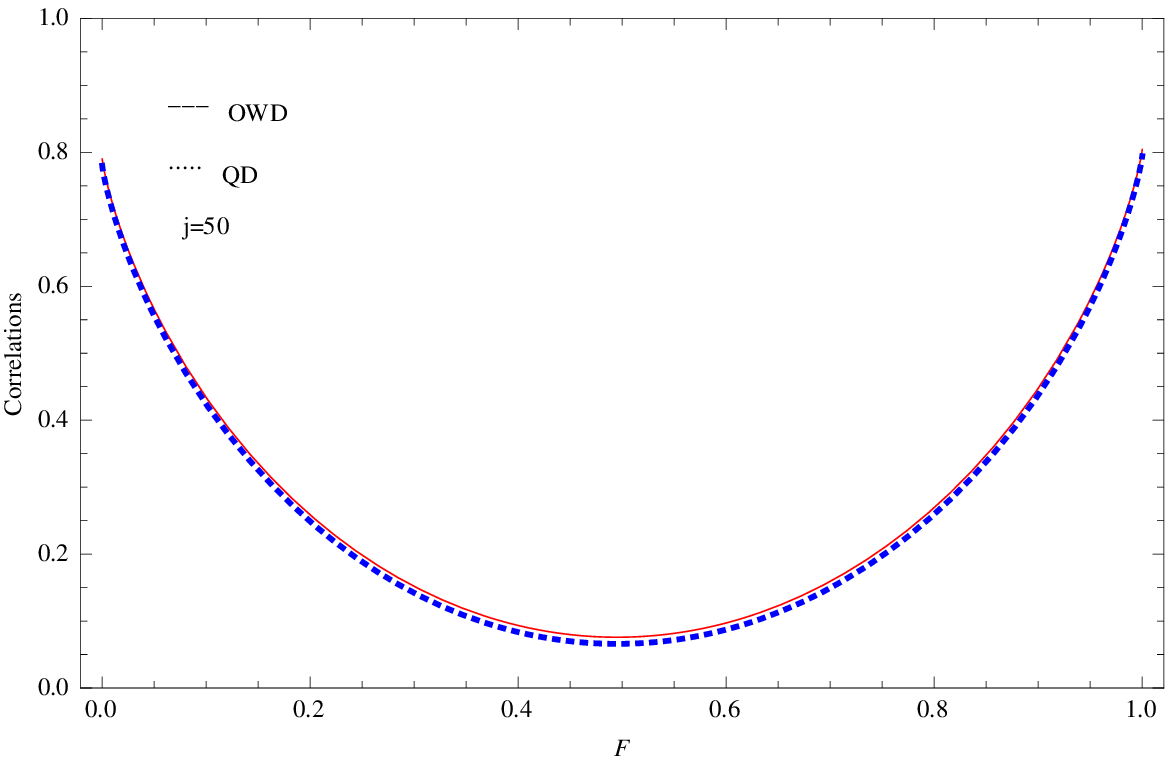}
\caption{(Color online) quantum correlations of the bipartite state
composed of a spin-$j$ and a  spin-$\frac{1}{2}$ vs $F$. (a) one-way
deficit of $j=1$(green line), $j=2$(yellow line), $j=3$(black line) and
$j=\frac{1}{2}$(red line), $j=\frac{3}{2}$(blue line), $j=\frac{5}{2}$(gray line),
(b) one-way deficit of $j=300$(red line) and
$j=\frac{599}{2}$(blue line), (c) comparison of one-way
deficit (solid line) and quantum discord(dotted line) of $j=1$(green line), $j=2$(black line )
and $j=\frac{1}{2}$(red line), $j=\frac{3}{2}$(blue line), (d)
comparison of one-way deficit( red solid line) and quantum discord (blue dotted line) of
$j=50$.}
\end{center}
\end{figure}

In Fig.1 we show that quantum correlations is a function of $F$ for different
spin $j$.  For half-integer $j$, we observe that as $j$
increases, the one-way deficit increases for small $F$ and decreases for
large $F$. But for integer $j$, the one-way deficit decreases as
$j$ increases, see Fig.1(a). For high
dimensional system (large $j$), the one-way deficit becomes
symmetric around the point $F=\frac{1}{2}$ where $\Delta^{\rightarrow}(\rho^{ab})$ vanishes, see Fig.1(b).
Maximum of the one-way deficit is attained at $F=1$  and at
$F=0$ for all dimensional systems.

From Eq.(\ref{deficit}), one can see an interesting fact: for
half-integer $j$, the one-way deficit is equal to the quantum discord, see Fig.1(c).
For integer $j$, the difference between the one-way deficit and the quantum discord
is $\frac{1}{d}$,  see Fig.1(c), and the difference tends to be small
for large $j$, see Fig.1(d).

We now compare the one-way deficit with the entanglement of formation (EoF).
The EoF for a spin-$\frac{1}{2}$ and a
spin-$j$ $SU(2)$ invariant state $\rho^{ab}$ is given by \cite{Breuer2},
\begin{eqnarray}
EoF(\rho^{ab})= \left\{ \begin{array}{ll}
        0, & F\in[0,\frac{2j}{2j+1}] \\[2mm]
        H\left(\frac{1}{2j+1}\left(\sqrt{F}-\sqrt{2j(1-F)}\right)^2 \right), & F\in[\frac{2j}{2j+1},1], \end{array} \right.
\end{eqnarray}
where $H(x)=-x\text{log}x-(1-x)\text{log}(1-x)$ is the binary
entropy. It is shown that EoF becomes a upper bound for
quantum discord in $d\otimes d$ Werner states \cite{Chitambar}.
However, we can see that the one-way deficit always remains larger than
EoF for half-integer $j$. The difference between EoF and $\Delta^{\rightarrow}(\rho^{ab})$
increases as $j\rightarrow\infty$ \cite{cakmak}. But for integer $j$, the difference
decreases as $j\rightarrow\infty$, see Fig.2 (a) and (b). It should be
noted that as $j\rightarrow\infty$, the state $\rho^{ab}$ becomes
separable while its one-way deficit remains finite.
In the region in of zero EoF, the one-way deficit survives.

\begin{figure}[h]
\raisebox{12em}{(a)}\includegraphics[width=6.25cm]{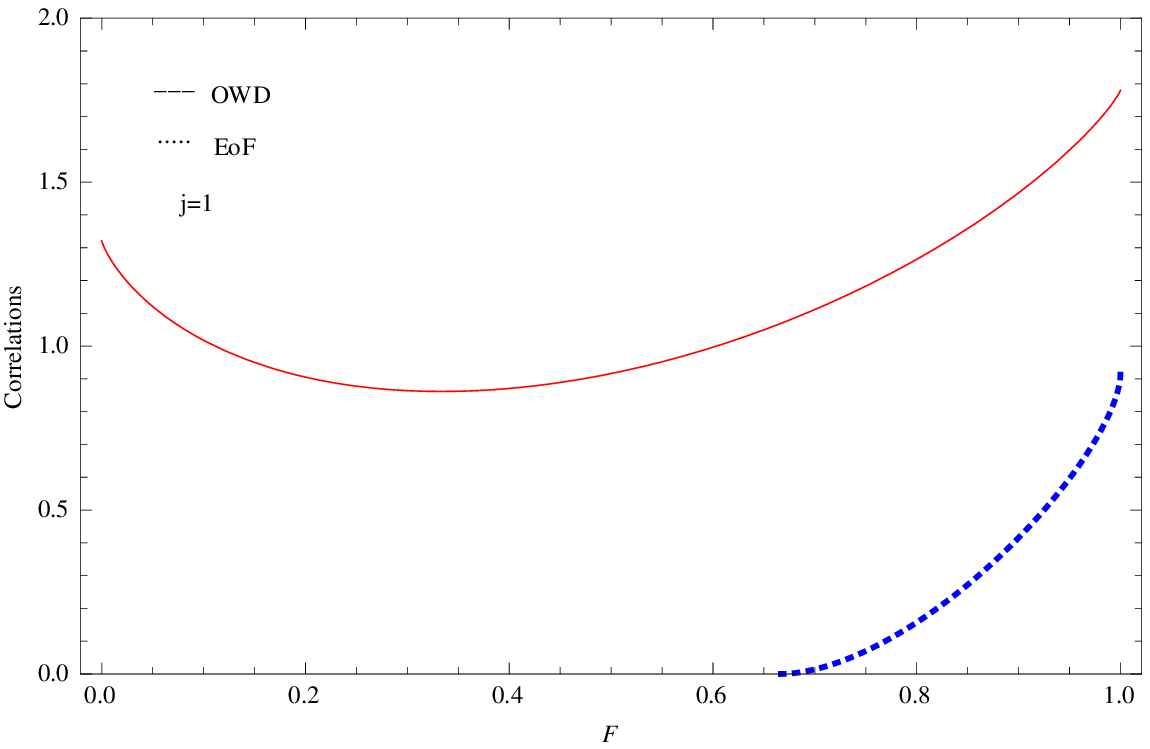}\qquad
\raisebox{12em}{(b)}\includegraphics[width=6.25cm]{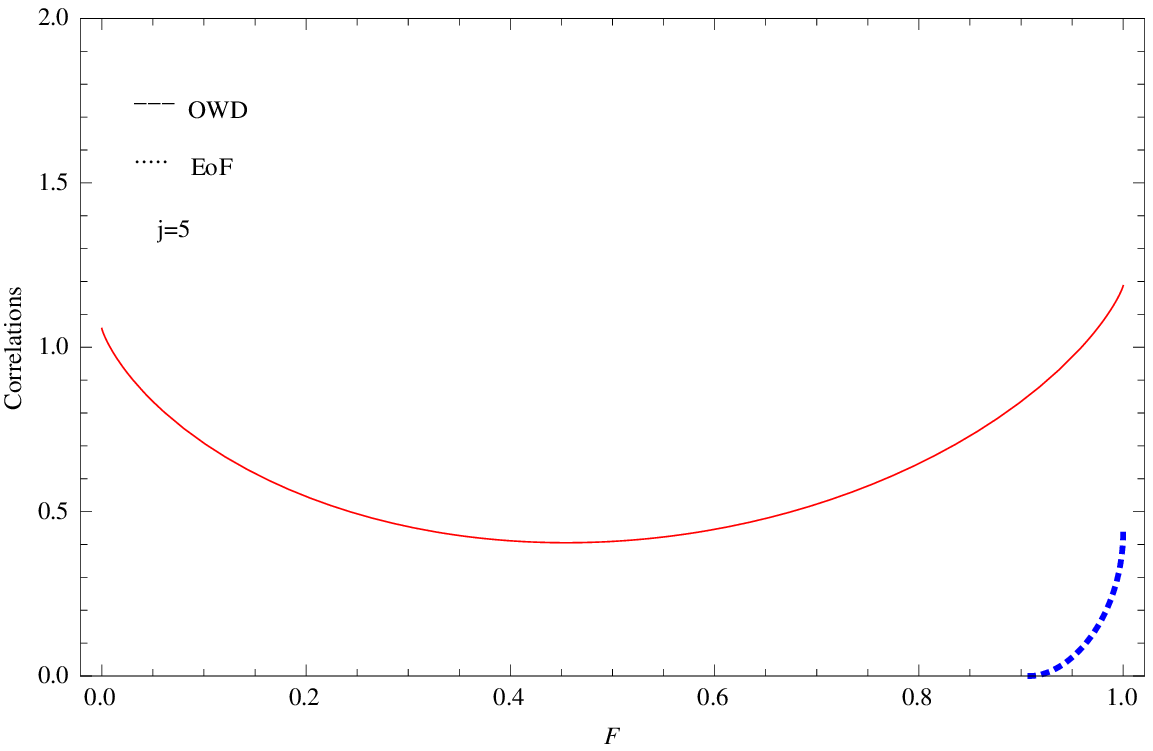}
\label{Fig:1}
\end{figure}
\begin{figure}[h]
\begin{center}
\caption{(Color online) one-way deficit (red solid line) and EoF (blue dotted line) vs. $F$: (a) $j=1$, (b) $j=5$.}
\end{center}
\end{figure}

\section{Conclusion}

We have analytically calculated the one-way deficit of $SU(2)$
invariant states consisting of a spin-$j$ and a spin-$\frac{1}{2}$
subsystems, with measurement on the spin-$\frac{1}{2}$ subsystem.
By comparing the one-way deficit with the quantum discord of
these states we have shown that the one-way deficit is equal to the quantum discord for
half-integer $j$, and the one-way deficit is larger than the quantum discord for
integer $j$. We have also compared our results on one-way deficit with the quantum
entanglement EoF. It is shown that in the large $j$, one-way deficit remains significantly larger than EoF.
Moreover, the maximal value of one-way deficit
decreases with the increasing system size. As there are abundance of $SU(2)$
invariant states in real physical systems, our results can be used
in quantum protocols that rely on one-way deficit.

\bigskip
\noindent{\bf Acknowledgments}\, \, This work is supported by
NSFC11275131 and KZ201210028032.

\end{document}